\documentclass[aps,twocolumn,showpacs,prl]{revtex4-1}
\usepackage{amsmath}
\usepackage{amssymb}
\usepackage{graphicx}
\usepackage{dcolumn}
\usepackage{bm}

\usepackage{bm}
\usepackage{graphicx}
\usepackage{ulem}

\def\be{\begin{equation}}
\def\ee{\end{equation}}
\def\ber{\begin{eqnarray}}
\def\eer{\end{eqnarray}}
\def\bwt{\begin{widetext}}
\def\ewt{\end{widetext}}

\def\e{{\varepsilon}}
\def\o{{\omega}}
\def\bt{\textbf}

\begin{document}

\draft
\title {Plasmon and dielectric background inhomogeneity enhancement of Coulomb drag\\ in graphene double-layer structures}
\author{S. M. Badalyan}
\email{Samvel.Badalyan@ua.ac.be}
\affiliation{Department of Physics, University of Antwerp, Groenenborgerlaan 171, B-2020 Antwerpen, Belgium}
\author{F. M. Peeters}
\affiliation{Department of Physics, University of Antwerp, Groenenborgerlaan 171, B-2020 Antwerpen, Belgium}

\date{\today}

\begin{abstract}
The drag of massless fermions in graphene double-layer structures is investigated in a wide rage of temperatures and inter-layer separations. We show that the inhomogeneity of the dielectric background in such graphene structures for experimentally relevant parameters results in a significant enhancement of the drag resistivity. At intermediate temperatures the dynamical screening via plasmon-mediated drag enhances the drag resistivity and results in an upturn in its behavior at large inter-layer separations. In a range of inter-layer separations, corresponding to the strong-to-weak crossover coupling of graphene layers, we find that the drag resistivity decreases  approximately quadratically with the inter-layer spacing. This dependence weakens with a decrease of the inter-layer spacing while for larger separations we recover the cubic (quartic) dependence at intermediate (low) temperatures.
\end{abstract}


\maketitle

\paragraph{Introduction}
The discovery of graphene \cite{Graphene,GeimNovo2007}, a monolayer lattice of carbon atoms, opened up possibilities for exploring new phenomena in fundamental physics \cite{Kotov,SarmaRMP2011} and for creating 
a new generation of electronic device applications \cite{Geim2009,GeimMac}. Based on monolayer graphene, novel double-layer structures have been recently realized experimentally \cite{Tutuc2011,Ponamarenko2011,Ponamarenko2012} where massless fermions in two layers are coupled only via many-body Coulomb interaction. Some current efforts in graphene physics focus on graphene double-layer structures (GDLS)  \cite{Tutuc2012,SMB2012,Profumo2012,Stauber2012,Profumo2010,Hwang2009,Vignale2012} with the aim to find new electronic properties, which will emerge from the inter-layer many-body Coulomb interaction of massless, chiral fermions with a unique Dirac-like spectrum.

Frictional drag \cite{Rojo1999} between spatially separate electron layers provides one of the most powerful tools for the study of interaction effects. Recently, Coulomb drag in GDLS has attracted substantial theoretical \cite{Matos2012,Polini2012,Narozhny2012,Hwang2011,Katsnelson2011,Castro2011,Narozhny2007,Tse2007} and experimental \cite{Tutuc2011} attention. Despite much efforts, no general agreement has been reached between experimental observations and theory, and there is still no clear understanding of the dependence of the drag on the inter-layer spacing and on the carrier density while the full treatment of the plasmon-mediated drag and of the dielectric inhomogeneity effect for realistic samples is still missing.

In the present paper we calculate the drag resistivity in GDLS in a wide rage of temperatures, $T$, and of inter-layer separations, $d$, using the {\it finite-T} polarizability and  the {\it finite-T} nonlinear susceptibility for individual graphene layers. We focus on three main questions. First, we investigate the effect of the dielectric inhomogeneity of the GDLS surrounding environment on the drag. Then, we calculate the contribution to the drag made by double-layer optical and acoustical plasmon modes and study the dependence of the drag rate on the inter-layer spacing. We show that Coulomb drag in GDLS immersed in a three layer nonhomogeneous dielectric medium (see Fig.~1(left)) is significantly larger than that calculated for the respective averaged homogenous background. This enhancement is observed for temperatures up to the Fermi temperature, $T_{F}$, and it becomes larger with increase of the inter-layer spacing. We find that at intermediate temperatures the dynamical screening of the inter-layer Coulomb interaction results in the plasmon enhancement of drag, which is strongly pronounced at large inter-layer separations. Our \begin{figure}[h]
\begin{center}
\begin{minipage}{.375\linewidth}
\centering
\includegraphics[width=\linewidth]{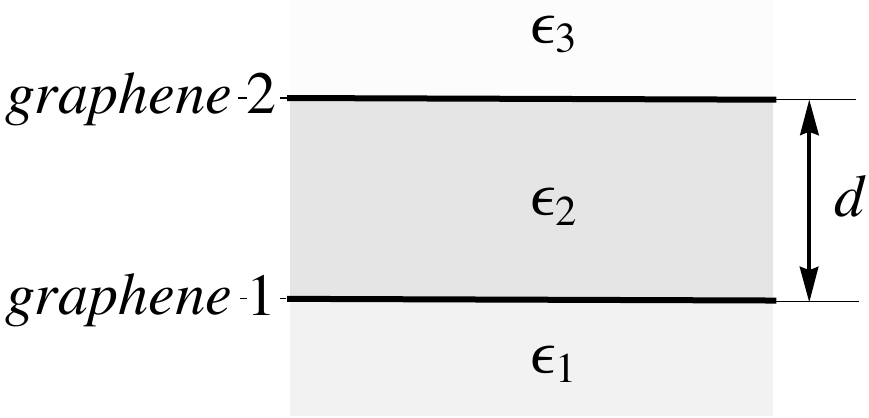}
\end{minipage}
\hfill
\begin{minipage}{.575\linewidth}
\centering
\includegraphics[width=\linewidth]{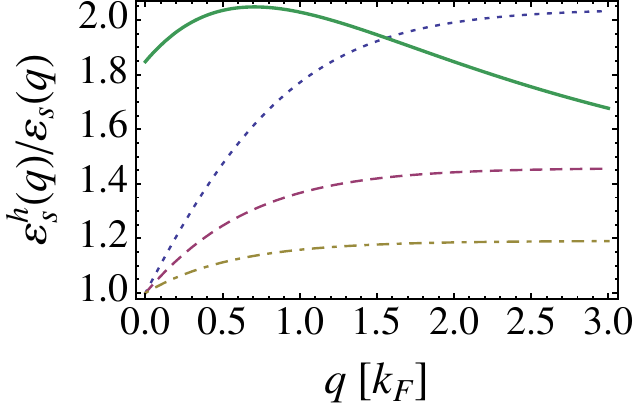}
\par\vspace{0pt}
\end{minipage}
\end{center}
\caption{
(left) A graphene double-layer system immersed in a three layer dielectric medium. The solid lines with spacing $d$ represent the active and passive graphene layers, $1$ and $2$, separating different materials with dielectric permittivities $\epsilon_{1}$, $\epsilon_{2}$, and $\epsilon_{3}$. 
(right) The solid curve corresponds to the ratio of the double-layer screening functions $\e^{h}_{s}(q)$ to $\e_{s}(q)$, which are calculated within the static screening approximation, respectively, in GDLS with a homogeneous  average dielectric permittivity $\bar{\epsilon}_{13}$ and in GDLS with a nonhomogeneous dielectric background, corresponding to the three layered medium of the left figure. The dotted, dashed and dot-dashed curves show, respectively, the intra- and inter-layer effective dielectric functions, $\bar{\epsilon}_{11}(q d )$, $\bar{\epsilon}_{22}(q d )$, and $\bar{\epsilon}_{12}(q d )$ in units of $\bar{\epsilon}_{13}$. The electron density in each graphene layer is $n_{1}=n_{2}=10^{12}$ cm$^{-2}$.
}
\label{fig1}
\end{figure}
calculations for $d=30$ nm  show an upturn in the drag resistivity at approximately $0.15T_{F}$. We calculate the drag rate both in the regime of strongly ($k_{F}d\lesssim 1$) and weakly ($k_{F}d\gg 1$) coupled graphene layers ($k_{F}$ is the Fermi wave vector). For inter-layer separations such that $k_{F}d\sim 1$ the decrease of the drag resistivity with $d$ is approximately quadratic and it weakens with increase of $d$. In the weakly coupled regime we recover the strong $d^{-4}$ dependence of the drag resistivity, calculated at $T=0.1T_{F}$ within the static screening approximation. At $T=0.2T_{F}$ the inclusion of the plasmon-mediated drag via the dynamical screening weakens the drag dependence on the spacing and results in a $d^{-3}$ behavior.

\paragraph{Theoretical concept}
Frictional Coulomb drag in double-layer electron systems manifests itself when an electrical current with density $J_{1}$ driven along the active layer induces, via momentum transfer due to inter-layer Coulomb interaction, an electric field $E_{2}$ in the passive layer, which is an open circuit. The transresistivity, defined as $\rho_{D}=-E_{2}/J_{1}$, is the direct measure of drag, which is studied in experiment. In terms of the diagonal intra-layer conductivities, $\sigma_{1}$ and $\sigma_{2}$, and the off-diagonal inter-layer drag conductivity, $\sigma_{D}$, the drag resistivity $\rho_{D}$ can be obtained by inverting the conductivity tensor. Assuming, according to the experimental situation, that $\sigma_{D}\ll \sigma_{1,2}$, we have $\rho_{D} \approx -\sigma_{D}/\sigma_{1}\sigma_{2}$. The drag conductivity between two graphene monolayers can be calculated from the general drag formula, derived within the Boltzmann equation \cite{Jauho1993}, the memory function \cite{Zheng}, and the Kubo \cite{Kamenev,Flensberg1} formalisms in the lowest order of perturbation theory with small inter-layer interaction. In GDLS the drag conductivity is represented as
\be\label{DC}
\sigma_{D}=\frac{1}{16\pi T \cal{A}}\sum_{\vec{q}}\int d\o \left| V_{12}(q,\o)\right|^{2} \frac{\Gamma_{1}(q,\o)\Gamma_{2}(q,\o)}{\sinh^{2}\o/2T}
\ee
where $\cal{A}$ is the normalization area, $\o$ and $q$ are the transferred energy and momentum from layer 1 to  layer 2 at temperature $T$. The dynamically screened Coulomb propagator $V_{12}(q,\o)$ describes the charge density fluctuations that realize the electron-electron interaction between the graphene layers. Within the random phase approximation $V_{12}(q,\o)$ can be obtained from a standard $2\times2$ matrix Dyson equation as
\be\label{ICI}
V_{12}(q,\o)= \frac{v_{12}(q)}{\e(q,\o)}
\ee
where the double-layer screening function is
\begin{equation}\label{Screen}
\varepsilon(q,\omega)=\varepsilon_{1}(q,\omega)\varepsilon_{2}(q,\omega)-v_{12}(q)^{2}\Pi^{0}_{1}(q,\omega)\Pi^{0}_{2}(q,\omega)
\end{equation}
with the screening function and the Lindhard polarization function \cite{GV2005} of graphene monolayers given, respectively, by $\varepsilon_{1,2}(q,\omega)=1-v_{11,22}(q)\Pi^{0}_{1,2}(q,\omega)$ and $\Pi^{0}_{1,2}(q,\omega)$. In the actual calculations of the drag resistivity we make use of the exact semi-analytical formulas from Ref.~\onlinecite{ramez} for the finite temperature polarizability $\Pi_{1,2}^{0}(q,\omega)$ of graphene monolayers, including both the  inter-chirality and the intra-chirality subband electronic transitions. The temperature dependent chemical potential $\mu(T)$ is determined by the carrier density $n$ from the equation $\text{Li}_2\left(-\exp(-\bar{\mu}(t)/t) \right)-\text{Li}_2\left(-\exp(\bar{\mu}(t) /t)\right)=n/2 t^2$ where $\text{Li}_2\left(x\right)$ is the dilogarithm function, $\bar{\mu}=\mu/T_{F}$ and $t=T/T_{F}$. In the static screening approximation we use the total Lindhard polarization function, $\Pi^{0}(q,0)=2k_{F}/\pi v$, in the static limit for small momenta $q\leq 2k_{F}$. Here $v$ is the velocity of the Dirac fermions.

In double-layer structures the bare Coulomb interaction is given by a $2\times2$ matrix, which, in general, represents three different electron-electron interactions, the intra-layer, $v_{11}(q)$ and $v_{22}(q)$, and the inter-layer, $v_{12}(q)=v_{21}(q)$, given by
\begin{eqnarray}\label{BCI}
v_{ij}(q,d)=\frac{2\pi e^{2}}{q \bar{\epsilon}_{ij}(q d)}
\end{eqnarray}
where $i, j = 1,2$ are the graphene layer indices. The effective dielectric functions, $\bar{\epsilon}_{ij}(q d)$, take into account the inhomogeneity of the dielectric background of the GDLS, which plays an important role in determining the bare Coulomb interactions. In the dielectric environment consisting of three contacting media with different, frequency independent dielectric permittivities $\e_{1}$, $\e_{2}$, and $\e_{3}$, the effective dielectric functions of GDLS are obtained from the exact solution of the Poisson equation for the Coulomb potential and are given by Eqs.~(5)-(7) of Ref.~\onlinecite{SMB2012}.
As seen in Fig.~1(right) the dielectric inhomogeneity modifies largely the behavior of the bare Coulomb interactions and, what is especially important, the behavior of the screening function in momentum space. In the long wavelength limit all three interactions are determined by the same effective dielectric constant, given by the arithmetic average of the top and bottom surrounding media, $\bar{\epsilon}_{13}=(\epsilon_{1}+\epsilon_{3})/2$, and does not depend on the dielectric constant, $\epsilon_{2}$, of the inter-layer spacer. This is not true for the double-layer screening function $\e(q,\o)$. It is seen in Fig.~1(right) that the static screening of GDLS $\e_{s}(q)=\e(q,0)$ in the long wavelength limit differs essentially from the screening function, $\e_{s}^{h}(q)$, calculated for GDLS, immersed in a homogeneous dielectric medium with an average permittivity $\bar{\epsilon}_{13}$. As we see below this effect results in an overall significant enhancement of the drag resistivity.

Furthermore, the other important quantity in Eq.~(\ref{DC}) is the quadratic response function of the charge density to an external potential, $\Gamma_{i}(q,\o)$. Due to the linear dispersion of Dirac fermions, $\Gamma_{i}(q,\o)$ is no longer proportional to the imaginary part of the individual layer polarizability as is the case in the weak scattering limit for the usual two-dimensional electron gas with parabolic dispersion. 
Here in the actual calculations of the drag resistivity we use the finite temperature nonlinear susceptibility for individual graphene layers from Ref.~\onlinecite{Narozhny2012} where the carrier transport time $\tau_{tr}$ is approximated by a constant. This approximation is well justified at low temperature $T\ll T_{F}$ \cite{Polini2012,Narozhny2012}. The energy dependence of the scattering time can be important in the vicinity of the Dirac point when $\mu(T)\ll T$ \cite{Narozhny2012}. Here we limit ourselves to the important range of not very high temperatures, $T\lesssim T_{F}$. Therefore, the use of the constant scattering time approximation near the upper limit of temperatures $T=T_{F}$ should be still justified. Here we assume also that the intra-layer conductivities are restricted by impurity scattering and use $\sigma_{1,2}=e^{2}\tau_{tr}\e_{1,2 F}\eta(t)/\pi$ where $\eta(t)=t \int^{\infty}_{-\infty} d z |z| / \cosh^2(z+\mu(t)/2t)$ for the temperature dependent intra-layer conductivities.

\begin{figure}[t]
\includegraphics[width=.9\linewidth]{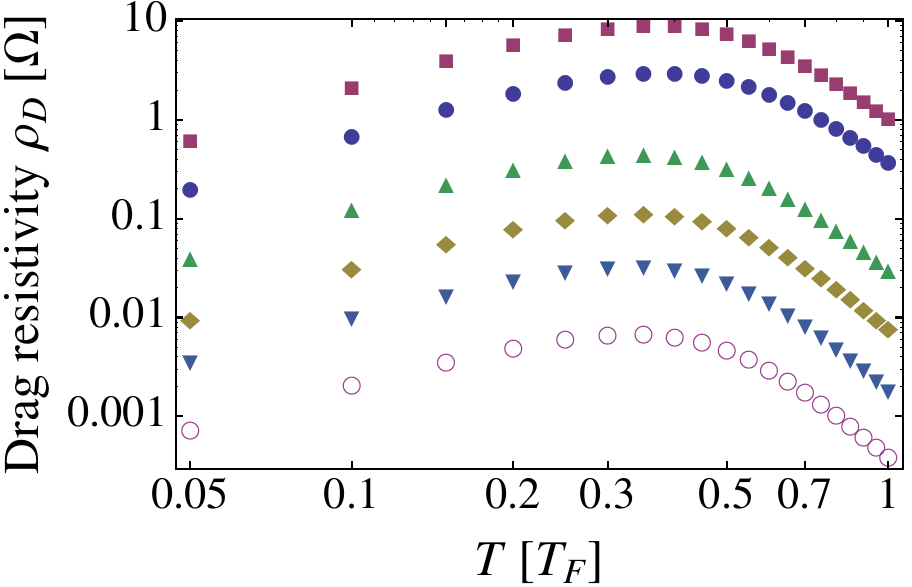}
\caption{
The effect of the dielectric background inhomogeneity on Coulomb drag in GDLS. The top pair of symbol sets shows the log-log plots of the drag resistivity as a function of scaled temperature in GDLS with  $d=5$ nm inter-layer spacing, immersed in a nonhomogeneous (the upper set) and homogeneous (the lower set) dielectric background with the parameters of GDLS of Fig.~\ref{fig1}. The two other pairs of data sets in the mid and bottom of the figure correspond, respectively, to calculations for the $d=15$ and $30$ nm spacing. All curves have been calculated  within the static screening approximation.}
\label{fig2}
\end{figure}

\paragraph{Results and discussions} Based on the formalism described above we present here our numerical calculations of the drag resistivity, carried out in a wide range of inter-layer separations and temperatures up to the Fermi temperature $T_{F}$. 
In Fig.~\ref{fig2} we study the effect of the dielectric background inhomogeneity on drag of massless fermions by comparing the drag resistivity calculated for GDLS immersed in a nonhomogeneous dielectric background with that calculated for GDLS in a homogeneous background. The presented results for three different values of the inter-layer spacing, $d=5$, 15, and 30 nm, show that the effect of the dielectric inhomogeneity is important in GDLS with realistic dielectric parameters, corresponding to the experimental samples of Ref.~\onlinecite{Tutuc2011}. This new effect comes from the momentum dispersion of the effective dielectric functions for the three layer nonhomogeneous dielectric medium (see Ref.~\onlinecite{SMB2012}). It changes the respective bare Coulomb interactions (\ref{BCI}) and reduces the screening function (\ref{Screen}) in a nonhomogeneous dielectric background ({\it cf.} Fig.~\ref{fig1}) that in its turn enhances the effective inter-layer Coulomb interaction given by Eq.~(\ref{ICI}). We find that for $d=5$ nm the drag resistivity in GDLS with a nonhomogenous dielectric background is larger by a factor of $3$ than that in a corresponding homogeneous dielectric medium.  This increase strengthens with the inter-layer spacing and for $d=30$ nm the difference is a factor of $4.7$. As seen in Fig.~\ref{fig2}, the overall qualitative temperature dependence of the drag resistivity is not affected when we include the effect of the dielectric inhomogeneity or when we vary the inter-layer spacing. At low temperatures the drag resistivity increases quadratically with $T$ due to the thermal broadening of the Fermi surface and at temperatures approximately $T=0.35T_{F}$ it shows a maximum. At higher temperatures the chemical potential $\mu(T)$ decreases and becomes smaller than $T$, the electron gases in the two graphene layers behave as a Boltzmann gas and the drag resistivity decreases approximately as the fourth power of temperature. 

\begin{figure}[t]
\includegraphics[width=.9\linewidth]{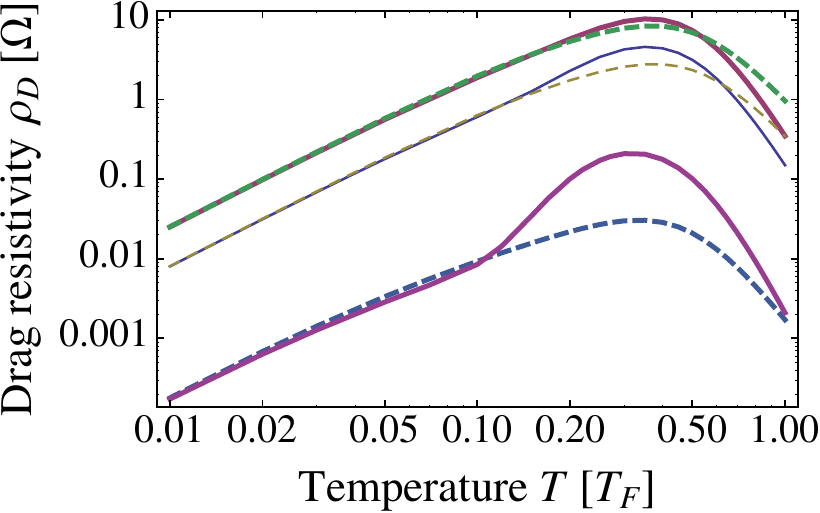}
\caption{Plasmon enhancement of Coulomb drag of massless fermions in GDLS. The top and bottom bold curves represent the log-log plots of the drag resistivity versus scaled temperature for the $d=5$ and $d=30$ nm inter-layer spacing, respectively.  The solid curves correspond to calculations using the finite temperature exact polarization and nonlinear response functions while the dashed curves are calculated within the static screening approximation. The parameters are the same as in Fig.~\ref{fig1} for GDLS with a nonhomogeneous dielectric background. The thin curves correspond to the calculations for $d=5$ nm in GDLS with a homogeneous dielectric background.
}
\label{fig3}
\end{figure}

\begin{figure}[t]
\includegraphics[width=.9\linewidth]{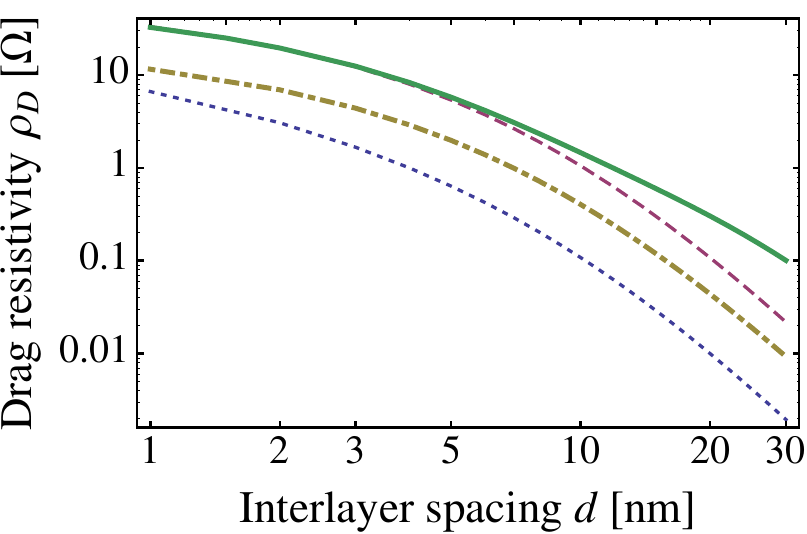}
\caption{The log-log plot of the drag resistivity as a function of the inter-layer spacing $d$ for $T=0.2T_{F}$ (the solid and dashed curves) and for $T=0.1T_{F}$ (the dot-dashed and dotted curves). The solid curve corresponds to calculations using the finite temperature exact polarization and nonlinear response functions, the other curves are based on the static screening approximation. The dotted curve is calculated for GDLS with a homogeneous dielectric background, the other curves for GDLS with a nonhomogeneous dielectric background. The parameters are the same as in Fig.~\ref{fig1}.}
\label{fig4}
\end{figure}

One can see in Fig.~\ref{fig3} that the dynamical screening changes the temperature dependence of the drag resistivity, especially, at intermediate and higher temperatures. For $d=5$ nm the static screening approximation provides an adequate description of drag at low temperatures up to $0.2T_{F}$. At intermediate temperatures the double-layer optical and acoustical plasmon modes become thermally excited and the drag resistivity increases a little by the plasmon-mediated drag in comparison with that obtained in the static approximation. At even higher temperatures, $T\sim T_{F}$, the dynamical screening has the opposite effect, it increases with $T$ due to the temperature dependence of the dynamical polarizability and therefore the decrease of the transresistivity with $T$ near $T_{F}$ becomes stronger ({\it cf.} the solid and dashed curves). For the larger inter-layer spacing of $d=30$ nm the static screening approximation slightly overestimates the drag rate at low temperatures. One can see, however, that the dynamical screening causes an upturn in the drag resistivity at approximately $T_{c}=0.15T_{F}$. With an increase of the inter-layer spacing the drag mediated by the electron-hole fluctuations decreases with $d$ much faster than the plasmon-mediated drag ({\it cf.} the solid and dashed curves in Fig.~\ref{fig4}). Therefore the plasmon enhancement is more strongly pronounced for $d=30$ nm than for $d=5$ nm. Note that the upturn temperature $T_{c}$ in GDLS is smaller than that obtained for the usual two-dimensional electron gases \cite{Flensberg2}.

In Fig.~\ref{fig4} we study the inter-layer spacing dependence of the drag resistivity at $T=0.1T_{F}$ and $T=0.2T_{F}$. This dependence is mainly determined by the parameter $k_{F}d$, which is a measure of the inter-layer coupling. For small values of $k_{F}d\lesssim 1$ (for the density $n$ considered here we have $k_{F}d\sim 1$ for $d\sim 6$) the typical momenta $q$ in drag scattering events are of the order of $q_{T}=T/v$ and do not depend on $d$. In this case we find that the drag resistivity decreases as $\rho_{D}\propto d^{-\delta}$ with $\delta\lesssim 2$. For larger separations with $k_{F}d\gg 1$ the drag mediated by the electron-hole fluctuations is dominated by the momenta $q\lesssim d^{-1}$ and we find that the drag resistivity calculated within the static screening approximation behaves approximately as $d^{-4}$ \cite{Jauho1993,SMB2007}. At large separations the effect of the dielectric inhomogeneity on the spacing dependence of drag is weak ({\it cf.} the dotted and dotdashed curves in Fig.~\ref{fig4} have almost the same dependence on $d$). However, the dynamical screening due to the double-layer plasmons weakens the inter-layer spacing dependence of the drag resistivity and we find that it behaves (the solid curve) approximately as $d^{-3}$.

In conclusion, we investigated the drag of massless fermions in GDLS in a wide range of temperatures and inter-layer separations. Our theory includes the effect of dielectric background inhomogeneity, which for realistic parameters results in a significant increase of the drag rate. At intermediate temperatures the thermally excited  double-layer plasmons cause an upturn in the drag resistivity. We find that the drag resistivity decreases with the inter-layer spacing approximately quadratically for inter-layer separations, corresponding to the strong-to-weak inter-layer coupling crossover. This dependence increases (decreases) with an increase (decrease) of the spacing.

We acknowledge support from the Flemisch Science Foundation (FWO-Vl) and the Belgian Science Policy (BELSPO).

\end{document}